\documentclass[twocolumn,showpacs,amsmath,amssymb]{revtex4}
\usepackage{graphicx}
\usepackage{dcolumn}
\usepackage{bm}
\usepackage{amsmath}
\usepackage{amsfonts}
\begin{document}

\title{On (Not)-Constraining Heavy Asymmetric Bosonic Dark Matter}
\author{Chris {\sc Kouvaris}}\email{kouvaris@cp3.sdu.dk}
\affiliation{$\text{CP}^3$-Origins, University of Southern Denmark}
\affiliation{Danish Institute for Advanced Study, Campusvej 55, Odense 5230, Denmark}
\author{Peter {\sc Tinyakov}}\email{Petr.Tiniakov@ulb.ac.be}
 \affiliation{Service de Physique Th\'eorique,  Universit\'e Libre de Bruxelles, 1050 Brussels, Belgium}
 
\begin{abstract}
Recently, constraints on bosonic asymmetric dark matter have been imposed
 based on the existence of old neutron stars excluding the dark matter
masses in the range from $\sim 2$~keV up to several GeV. The constraints are
based on the star destruction scenario where the dark matter particles captured by the
star collapse forming a black hole that eventually consumes the host star. In
addition, there were claims in the literature that similar constraints can be
obtained for dark matter masses heavier than a few TeV. Here we argue that it is not
possible to extend to these constraints. We show that in the case of heavy dark matter,
instead of forming a single large black hole that consumes the star, the
collapsing dark matter particles form a series of small black holes that evaporate fast
without leading to the destruction of the star. Thus, no constraints arise for
bosonic asymmetric dark matter particles with masses of a few TeV or higher.
 \\[.1cm]
 {\footnotesize  \it Preprint: CP$^3$-Origins-2012-034 \& DIAS-2012-35.}
\end{abstract}


\maketitle 

The last few years stellar observations have been used in order to constrain
specific dark matter candidates or predict interesting phenomena related to
dark
matter~\cite{Casanellas:2009dp,PerezGarcia:2010ap,Casanellas:2010he,Lopes:2011rx,Casanellas:2011qh,Casanellas:2011kf,PerezGarcia:2011hh,PerezGarcia:2011jb,Brayeur:2011yw,Iocco:2012wk,Horowitz:2012jd,Capela:2012jz,Lopes:2012af}.
Specifically, compact stars such as white dwarfs and neutron stars have been
found to impose severe constraints on some dark matter
models~\cite{Goldman:1989nd,Kouvaris:2007ay,Bertone:2007ae,Sandin:2008db,McCullough:2010ai,Kouvaris:2010vv,deLavallaz:2010wp,Kouvaris:2010jy,Ciarcelluti:2010ji,Kouvaris:2011fi,McDermott:2011jp,Kouvaris:2011gb,Guver:2012ba,Fan:2012qy}.
In principle, there are two types of effects that can take place in compact
stars and can give rise to constraints on dark matter.  The first type is
related to the thermal evolution of compact
stars~\cite{Kouvaris:2007ay,Bertone:2007ae,Kouvaris:2010vv,deLavallaz:2010wp}.
In this case, annihilation of trapped weakly interacting massive particles
(WIMPs) inside a compact star can produce significant amount of heat that can
change the thermal evolution of the star at later times. As a result, stars
old enough to be quite cold might maintain higher temperature due to the
released heat.  

The second type of constraints is related to asymmetric dark
matter~\cite{Goldman:1989nd,Kouvaris:2010jy,Kouvaris:2011fi,McDermott:2011jp,Kouvaris:2011gb,Guver:2012ba,Fan:2012qy}.
In this case WIMPs carry a conserved quantum number and there is an asymmetry
between the populations of WIMPs and anti-WIMPs, so that the annihilation is
impossible in the present-day universe where only the WIMPs remain. Such kind
of WIMPs can accumulate in a compact star quite efficiently as long as the
WIMP-nucleon cross section exceeds a certain critical value which in the case
of a neutron star is quite small ($\sigma \sim
10^{-46}~\text{cm}^2$)~\cite{Kouvaris:2007ay}, several orders of magnitude
smaller than current limits from direct searches. Under certain circumstances,
the amount of WIMPs accumulated during the lifetime of the star is sufficient
to result in a gravitational collapse of the the WIMPs into a black hole. In
the case of fermionic dark matter the amount of WIMPs needed for collapse is
$\sim M_{\text{Pl}}^3/m^2$, $M_{\text{pl}}$ being the Planck mass and $m$ the
mass of the WIMP. Unless one assumes very heavy WIMPs and/or extremely high
background dark matter density at the location of a neutron star, this amount of
WIMPs is difficult to accumulate during the lifetime of the star (although the
situation may change in the presence of attractive self-interactions~\cite{Kouvaris:2011gb}).

On the contrary, if WIMPs are of bosonic nature, there is no Fermi pressure
and the amount of WIMPs needed for gravitational collapse is significantly
lower~\cite{Mielke:2000mh}, 
\begin{equation}
M > M_{\rm min} 
= {2M_{\text{Pl}}^2\over \pi m}
= 9.5\times 10^{33}{\rm GeV} \left({m\over10 {\rm TeV}}\right)^{-1}.
\label{eq:Mcrit}
\end{equation}
This amount of WIMPs can be easily accumulated even by nearby known old
neutron stars with the standard assumption about dark matter density near the
Earth. As it was pointed out in~\cite{Kouvaris:2011fi}, once $\sim 10^{36}$
WIMPs have been accreted and thermalized within the neutron star, a
Bose-Einstein condensate (BEC) forms and all newly accreted WIMPs fall into
the ground state. This state is very compact, so the WIMPs in the condensate
start to self-gravitate way before the condition of Eq.~(\ref{eq:Mcrit}) is
satisfied. Thus, once the amount of WIMPs in the condensate meets the
criterion for the gravitational collapse (Eq.~(\ref{eq:Mcrit})), they form a black
hole. Note that only a fraction of all WIMPs (namely, those in the BEC) goes
into a black hole. If the resulting black hole starts to grow and can destroy
the host star in a reasonable time, constraints on dark matter arise as
follows from the well known existence of old (more than a few billion years
old) neutron stars.

The fate of the black hole inside the neutron star is determined by the two
competing processes: the accretion of surrounding nuclear matter that
increases the mass of the black hole and scales as $M_{\rm BH}^2$ ($M_{\rm
  BH}$ being the black hole mass), and Hawking radiation that scales as
$M_{\rm BH}^{-2}$. Therefore, there exists a critical value $M_{\text{crit}}$
such that heavier black holes grow until the whole star is consumed, while
lighter black holes evaporate completely. It follows from Eq.~(\ref{eq:Mcrit})
that the mass of the black hole formed by this mechanism is inversely
proportional to the WIMP mass, so that any black hole formed from WIMPs
heavier than a few GeV is lighter than $M_{\text{crit}}$ and evaporates
completely, in which case no constraints arise. 

In Refs.~\cite{McDermott:2011jp,Guver:2012ba,Fan:2012qy} it was
suggested that for WIMPs heavier than a few TeV the above mechanism is
modified and the constraints are recovered. The argument makes use of
the observation that for sufficiently heavy WIMPs, the
self-gravitation of the WIMP sphere sets in  before the formation
of BEC. At this point the  total amount of accumulated WIMPs
exceeds by many orders the amount of Eq.~(\ref{eq:Mcrit}) required for the black hole
formation, as well as the critical value $M_{\text{crit}}$. So if one
assumes that the collapse happens to the whole WIMP sphere at once the
resulting black hole is heavy enough to grow and destroy the
star. Hence the constraints reappear.

In this paper we demonstrate that this is not what happens. Although
it is true that for heavy WIMPs the self-gravitation sets in without
BEC formation, the collapse of the self-gravitating WIMP sphere is
hampered by the released gravitational energy and happens on a
timescale set by the cooling of the WIMPs. As the WIMP sphere
gradually shrinks, its density increases and BEC forms on a timescale
much shorter than the cooling time. From this stage on, the original
scenario is reproduced: the BEC region grows, becomes self-gravitating
and collapses into a black hole as soon as Eq.~(\ref{eq:Mcrit}) is
satisfied. The resulting black hole is too small to grow. Instead, it
evaporates on a time scale much shorter than it takes for the rest of
the WIMP sphere to collapse. The net result is the formation of tiny
black holes one after the other which evaporate shortly after their
birth without producing a heavy stable black hole. Thus no constraints
on heavy WIMPs arise.

Let us consider this argument more quantitatively. The WIMPs become
self-gravitating when their density becomes larger than that of the
nuclear matter in the core of the neutron star. This happens when the
total mass of the WIMPs exceeds 
\[
M_{\rm sg} = {4\pi\over 3} r_{\rm th}^3 \rho_c = 
\]
\begin{equation}
2.2\times 10^{40} {\rm GeV} 
\left({10 {\rm TeV}\over m}\right)^{3/2}
\left({T\over 10^5{\rm K}}\right)^{3/2},
\label{eq:self-grav}
\end{equation} 
where 
\begin{equation}
r_{\rm th} \simeq 2.2 ~\text{cm} 
\left( {T_c\over 10^5 {\rm K}} \right)^{1/2}
\left( {m\over 10 {\rm TeV}} \right)^{-1/2},
\label{eq:rth}
\end{equation}
is the thermal radius of the WIMP sphere inside the neutron
star at temperature $T$ and $\rho_c=5\times 10^{38}$~GeV/cm$^3$ is the star
core density. The density of WIMPs required to form BEC is~\cite{Kouvaris:2011fi} 
\begin{equation}
\rho_{\rm crit}\sim 4.7\times 10^{38}{\rm GeV\,
  cm}^{-3} \left({m\over10 {\rm TeV}}\right)^{5/2} \left ( \frac{T}{10^5\text{K}} \right )^{3/2},
\label{eq:BECdensity}
\end{equation}
which implies that the total WIMP mass necessary for the formation of BEC is
\begin{equation}
M_{\rm BEC} = 2.1\times 10^{40}{\rm GeV} 
\left({m \over10 {\rm TeV}}\right)
\left({T\over 10^5{\rm K}}\right)^{3}.
\label{eq:BECmass}
\end{equation}
From the dependence of Eqs.~(\ref{eq:self-grav}) and
(\ref{eq:BECmass}) on the WIMP mass $m$ one can see that for $m
\gtrsim 10$~TeV the self-gravitation sets in prior to the formation of
BEC. Note that the total mass of the WIMPs in this case is large
enough to overcome the constraint imposed by the uncertainty principle
of Eq.~(\ref{eq:Mcrit}).

Once the self self-gravitation sets in, the WIMP sphere starts to
collapse. However, it cannot collapse directly into a black hole because the
WIMPs have to lose their energy and momentum. Instead, the collapse
of WIMPs is likely to resemble the formation of dark matter halos, with the
essential difference that constant interactions with nucleons provide
an extra energy loss mechanism for the WIMPs. 
One may expect, therefore, that the WIMPs develop a cuspy
profile similar to the dark matter halos, which shrinks as the WIMPs lose
their energy in interactions with nucleons.

Even without the cusp, the shrinking WIMP sphere would form BEC long
before it reaches the size comparable to its Schwarzschild radius. To
make the argument as robust as possible, consider the worst
(unrealistic) case where the WIMP sphere contracts maintaining a
uniform density. In order to form a black hole of mass $M$ the WIMPs
have to reach the density $\rho_{\rm BH} \sim 3 (32\pi G^3 M^2)^{-1}
\sim 10^{74}~\mbox{GeV/cm}^3 (M/10^{40}{\rm GeV})^{-2}$, while the
density required for BEC formation is much lower,
c.f. Eq.~(\ref{eq:BECdensity}). Note that the smaller is the black
hole, the higher density is required, so the uniform contraction is indeed
the worst case. Thus, BEC will unavoidably be formed
before the WIMPs collapse into a black hole.

Consider now the formation of BEC in some more detail. Once the 
density of Eq.~(\ref{eq:BECdensity}) is reached, the formation of BEC happens on time
scales of order \cite{Stoof:1992zz}
\begin{equation}
t_{\rm BEC} \sim \frac{\hbar}{k_BT}  \sim10^{-16}{\rm s},
\end{equation}
i.e. practically instantaneously. Further shrinking of the WIMP
sphere results in increasing the mass of the condensate rather than
the density of non-condensed WIMPs. 

The size of the condensate can be estimated by noting that once the
WIMPs become self-gravitating they dominate the gravitational
potential. Taking their density approximately constant and equal to
that required for the condensate formation gives
\begin{eqnarray}
&r_{\rm BEC} &= \left( {8\pi\over 3} G\rho_{\rm crit} m^2\right)^{-1/4} \nonumber \\
&=& 1.7\times 10^{-6} {\rm cm} \left({m\over10 {\rm TeV}}\right)^{-9/8} \left ( \frac{T}{10^5\text{K}} \right )^{-3/8}.
\end{eqnarray}
This is much smaller than the size of the WIMP sphere. 

As the mass of the BEC grows, it eventually becomes self-gravitating
itself. This happens when the density of the condensate becomes equal
to the non-condensed WIMP density, i.e., when its mass exceeds
\begin{equation}
M_{\rm BEC,\,sg} = {4\pi\over 3} \rho_{\rm crit} r_{\rm BEC}^3 
= 9.6 \times 10^{21}{\rm GeV} \left({m\over10 {\rm TeV}}\right)^{-7/8}.
\end{equation}
Although self-gravitating, the condensate cannot yet collapse because
it does not satisfy the constraint of Eq.~(\ref{eq:Mcrit}). The latter is
satisfied when the total mass of the condensed state reaches $M_{\rm
  min}$.  Beyond that point stable configurations of the
self-gravitating condensate do not exist and it collapses into a black hole
\cite{Mielke:2000mh}. The resulting mass of the black hole is
 given by Eq.~(\ref{eq:Mcrit}). Note that this mass is much smaller
than the total mass of the WIMP-sphere which is of order $M_{\rm sg}$,
Eq.~(\ref{eq:self-grav}). Note also that the black hole mass becomes smaller
as the mass of the dark matter particle, $m$, increases.

The black hole of mass $M_{\rm min} $ is too small  to survive the Hawking
radiation. It evaporated on the time scale 
\begin{equation}
\tau = 5\times 10^3 {\rm s} \left({10 {\rm TeV}\over m }\right)^3. \label{evap}
\end{equation}

To conclude the argument, let us show that such burning of dark matter by the
formation and evaporation of small black holes is more efficient than
both the accretion of new dark matter onto the neutron star and the creation of
new black holes by the above mechanism. To this end consider the
relevant time scales. The accumulation of WIMPs by the neutron star
proceeds at a constant rate $F$ that depends on the local dark matter density
$\rho_{\rm dm}$ and equals~\cite{Kouvaris:2010vv} 
\begin{equation}
F = 1.25\times 10^{29}{\rm GeV/s} 
\left({ \rho_{\rm dm} \over 10^3 {\rm GeV/cm^3}}\right). 
\label{accretion}
\end{equation}
The time needed to accumulate the amount of dark matter equal to $M_{\rm  min}$
is 
\begin{equation}
t_{\rm acc} = 7.6\times 10^4 {\rm s} \left({m\over10 {\rm
    TeV}}\right)^{-1}
\left({ \rho_{\rm dm} \over 10^3 {\rm GeV/cm^3}}\right)^{-1}.
\end{equation}
This is larger by an order of magnitude than the evaporation time even for
$m=10$~TeV and even considering a $10^3$ GeV$/\text{cm}^3$ local dark matter
density. For nearby neutron stars where the dark matter constraints are more
reliable, this will give 5 orders of magnitude higher for 10 TeV WIMP mass.

Consider now the time needed to form a new black hole in the course of collapse of the
self-gravitating WIMP-sphere. According to the picture described above, the
contracting WIMP-sphere reaches at some point the density $\rho_{\rm crit}$ at
which the formation of the BEC becomes possible. Further contraction happens
at a constant WIMP density $\rho = \rho_{\rm crit}$ as the excess of the WIMPs
goes into the BEC state. To form a black hole, the amount of WIMPs equal $M_{\rm min}$
has to be moved from the WIMP-sphere to the BEC. The corresponding release of
potential energy is equal to the black hole mass times the difference of the
gravitational potential between the surface and the center of the sphere, 
\begin{equation}
\delta E = {1\over 2} {GMM_{\rm min} \over r_0}, 
\label{eq:energygain}
\end{equation}
where $M$ is the total WIMP mass and $r_0$ is the radius of the WIMP-sphere, 
\begin{equation}
r_0= \left ( \frac{3M_{\text{sg}}}{4 \pi \rho_{\text{crit}}} \right)^{1/3}
=2.2\, \text{cm} \left (\frac{10 \text{TeV}}{m} \right )^{4/3} . 
\label{eq:r0}
\end{equation}
The released energy has to be dispersed via collisions of WIMPs with nucleons,
which sets the time scale of the process.
Taking into account the Pauli blocking factor, the WIMP-nucleon collision
time is
\begin{equation}
\tau_{\rm col} = {1\over n \sigma v}\frac{4 p_F}{3m_N v}
= {2 p_F m\over 3\rho_c\sigma \epsilon},
\label{eq:tau_collision}
\end{equation}
where $p_F$ is the Fermi momentum of the nucleons, $m_N\simeq 1$~GeV is the
nucleon mass, $v$ is the WIMP velocity and $\epsilon$ is the WIMP kinetic energy
estimated as
\begin{equation}
\epsilon \sim {GMm\over 2 r_0} 
\label{eq:WIMPenergy}
\end{equation}
from the virial theorem. 
Taking into account that the typical energy loss per collision is $\delta
\epsilon = 2m_N\epsilon/m$ and assembling all the factors, the time required
to lose the excessive energy and form a black hole is
\begin{equation}
t_{\rm cool} = \tau_{\rm col} {m \delta E \over M \delta\epsilon } 
= {4 \over 3\pi}{ p_F\over m_N} 
{ r_0 M_{\rm Pl}^4 \over \rho_c \sigma M^2} .
\label{eq:t_cool}
\end{equation}
One can see that this time is shorter for larger 
mass $M$.  Substituting $M=M_{\rm sg}$ into Eq.~(\ref{eq:t_cool})
one gets 
\[
t_{\rm cool} \simeq 
1.5\times 10^3 {\rm s} 
\]
\begin{equation}
\times
\left({m\over10 {\rm  TeV}}\right)^{5/3}
\left({T\over 10^5 {\rm  K}}\right)^{-3}
\left({\sigma\over 10^{-43}\,{\rm cm}^2}\right)^{-1}.
\label{eq:t_cool_fin}
\end{equation}
This time is shorter by a factor of a few than the black hole evaporation time of
Eq.~(\ref{evap}). Note, however, the strong dependence of both quantities on
the WIMP mass $m$. Already for masses $m\gtrsim 13$~TeV  the black hole evaporation
time becomes shorter. 

Three comments are in order. The first comment is related to the black
hole evaporation. The fate of a black hole of mass $M_{\rm BH}$ is dictated by the equation
\begin{equation}
\frac{dM_{\rm BH} }{dt}=CM_{\rm BH}^2-\frac{f}{M_{\rm BH}^2},
\end{equation}
where the first term in the left hand side corresponds to the Bondi
accretion, and the second term corresponds to the Hawking
radiation. Here $C=4 \pi \lambda \rho_c G^2/c_s^3$,
$\lambda$ being a constant of order one and $c_s$ the speed of
sound. As we mentioned earlier, using Eq.~(\ref{eq:Mcrit}) one can
find that for WIMP masses above roughly 10~GeV, Hawking radiation wins
over accretion and the evaporation time is given by
Eq.~(\ref{evap}). However, the newly formed black hole can also
accrete from the dark matter population that, in the case examined
here, can have densities significantly higher than the nuclear matter
density 
$\rho_c$, as it can be seen from Eq.~(\ref{eq:BECdensity}). The
accretion of non-interacting collisionless particles  in the
non-relativistic limit onto a black hole is given by~\cite{Nov}
\begin{equation}
F=\frac{16 \pi G^2 M_{\rm BH}^2 \rho_{\rm dm}}{v_\infty},
\end{equation}
where $\rho_{\rm dm}$ is $\rho_{\rm crit}$ of
Eq.~(\ref{eq:BECdensity}), and $v_{\infty}$ is the average WIMP
velocity far away from the black hole. Using the virial theorem, we
take $v_\infty= \sqrt{GM/r}$, $M$ being the total mass of the WIMP
sphere and $r$ the radius of the WIMP sphere. It is understood that
for the first black hole $M=M_{\rm sg}$, and $r=r_0$. We have checked
that for WIMP sphere masses ranging from Eq.~(\ref{eq:Mcrit}) to
Eq.~(\ref{eq:self-grav}) the Hawking radiation dominates
overwhelmingly over the WIMP accretion despite the large WIMP density, and
therefore the black hole evaporation time is given accurately by
Eq.~(\ref{evap}).

The second comment is that the
cooling time derived above refers to the formation of the first black
hole of mass given by Eq.~(\ref{eq:Mcrit}). Subsequent black holes
require progressively longer times. This is easily seen from
Eq.~(\ref{eq:t_cool}) because $t_{\rm cool}$ scales as $1/M^{5/3}$
(recall that $r_0$ scales as $M^{1/3}$ from Eq.~(\ref{eq:r0})). Thus,
the more black holes have formed and evaporated, the smaller is the remaining
WIMP mass $M$ and the longer the time needed to form the next one.  

The third comment has to do with the neutron star temperature. As one can see
in Eq.~(\ref{eq:t_cool_fin}), there is a strong dependence in $T$. So
it is essential that $10^5$ K is a good estimate for the temperature
of the neutron star at the time the star has accumulated $M_{\rm sg}$
mass of WIMPs. Note that $M_{\rm sg}$ depends on the temperature,
cf. Eq.~(\ref{eq:self-grav}). The neutron star does not always has the
same temperature, but becomes colder with time due to different
cooling mechanisms. The cooling curves (i.e. the temperature of the star
as a function of the age) for typical neutron stars can be found,
e.g. in~\cite{Yakovlev:2004yr} and \cite{Kouvaris:2007ay}. Knowing
the cooling curves, $M_{\rm sg}(T)$ and the WIMP accretion rate from
Eq.~(\ref{accretion}), one can easily estimate the time it takes for the
star to accumulate a total WIMP mass of $M_{\rm sg}$. We found that in
all cases of interest (i.e. for $m$ ranging from 10 to 1000 TeV and
$\rho_{\rm dm}$ from 0.3 to $10^3$ $\rm GeV/\rm cm^3$), the
accumulation time is at least $\sim 1.5 \times 10^7$ years. By that
time the temperature of the star should be below $10^5$~K, which
justifies the use of this temperature in our estimates. 
 
Finally, let us examine one more factor that can potentially affect
our estimates. As the first black hole evaporates, it emits particles
via Hawking radiation that reheat the core of the star. In principle
such a deposit of heat can increase the temperature of nucleons and
WIMPs and thus could potentially change our estimates. We show here that
the increase in the temperature is small and the whole effect is
negligible.

The energy produced by the black hole evaporation, once transferred to
nucleons, propagates according to the diffusion equation,
 \begin{equation}
 \frac{\partial u}{\partial t} - D\nabla^2 u=Q,
 \end{equation}
where $u$ is the energy density due to the black hole evaporation, $D$ is the
diffusion coefficient and $Q$ is the energy injection rate.

Since the black hole is tiny and the mean free path of the Hawking-emitted
particles is short ($\sim 10^{-13}\,{\rm cm}$ assuming a typical
nucleon cross section $\sigma_N \sim \pi 10^{-26}~\rm cm^2$), one may
take $Q$ to be a delta-function in space, $Q=Q(t)\delta(\vec x)$.  We
will also assume that the energy is injected at a constant rate. As we
will see below, the time scale associated with the heat conductivity
of nuclear matter is extremely short, so a constant rate is a good
approximation for most stages of the black hole evaporation.

The diffusion coefficient $D=\kappa/c_V$ can be expressed in terms of
the specific heat capacity $c_V=1.4\times 10^{-9}\text{GeV}^3 (T/10^5\,{\rm
  K})$ of the nuclear matter \cite{Lattimer:1994} and the thermal
conductivity $k$. The latter can also be found in 
Ref.~\cite{Lattimer:1994}, $k \sim 10^3\,{\rm GeV}^2$.

Assuming that the energy release by the black hole starts at $t=0$, the
solution to the above diffusion equation is
\begin{equation}
  u(r,t)=\frac{Q}{4\pi Dr}\text{erfc} \left ( \frac{r}{\sqrt{4Dt}}
  \right ). 
\label{diffsol}
\end{equation}
The time-dependent factor saturates to one as soon as $r/\sqrt{4Dt}
\ll 1$ and the solution of Eq.~(\ref{diffsol}) becomes stationary at
corresponding distances. The characteristic time scale for reaching
the stationary solution  at distances of order of the size of the WIMP
cloud is
\begin{equation} 
t_\text{st}=\frac{r^2}{4D}\simeq 2.9 \times 10^{-9} \text{s}
\left (\frac{10\text{TeV}}{m} \right )^{8/3} \left (\frac{T}{10^5 \rm K} \right ) . 
\label{D}
\end{equation}
This time scale is much shorter than $t_{\rm cool}$ for all cases of
interest. The same time scale determines the onset of the equilibrium
once the heat source is switched off (the black hole has evaporated).

When the stationary solution is reached, the temperature of the
nuclear matter is given by  
\begin{equation}
T=\frac{Q}{4\pi kr}+T_{\infty},
\end{equation}
where $T_{\infty}=10^5$~K. The first term describes the effect
of the evaporating black hole. 

At distances comparable to the size of the WIMP-sphere $r\sim r_0$,
the change in the temperature during the black hole evaporation is $\sim
10~{\rm K}\, (m/10\,{\rm TeV})^{2}$, where we have estimated $Q$ by
dividing the black hole mass by its evaporation time. However, in cases where $t_{\rm cool}$ is longer than the
evaporation time, it is more appropriate to estimate the value of $Q$ by dividing the mass 
of the black hole by the $t_{\rm cool}$  of Eq.~(\ref{eq:t_cool_fin}). In this case, we find an increase in the temperature at most
\begin{equation}
\delta T \lesssim 35 {\rm K} 
\left (m\over 10~{\rm TeV} \right )^{-8/3}.
\label{eq:dTmax}
\end{equation}
Thus, the effect of the black hole evaporation on the temperature  in
the center of the neutron star is completely negligible.

\section{Conclusions}
We demonstrated that it is not possible to obtain constraints on
asymmetric bosonic dark matter with masses in the TeV range or higher
from the collapse of WIMPs into black holes inside neutron stars and
destruction of the latter. Although the self-gravitation of the
accumulated WIMPs starts before the BEC formation, we showed that the
WIMP sphere unavoidably  passes through the stage of the BEC formation as it
contracts. As a result, black holes with masses much smaller than the
total WIMP mass form and evaporate in times shorter
than both the accretion time scale and the WIMP sphere cooling time
scale.
This means
that the WIMP sphere does not collapse to a single large black hole
which would grow and consume the host star,
but instead collapses piece by piece in such a way that every black hole
that forms evaporates before another gets formed. The overall effect
is the (unobservable) heating of the neutron star, but not its
destruction. Thus no constraints on the TeV WIMP mass range can be
imposed.

We would like to stress that the  scenario we have studied is the most
conservative one. We assumed a uniform constant density for the WIMP
sphere. In reality, one should expect a more cuspy profile towards the
center of the WIMP sphere. As a result, $\rho_{\rm BEC}$ will be
achieved earlier, and the cooling time for formation of subsequent
black holes can increase substantially because the earlier the BEC
forms, the smaller will be the energy loss of WIMPs colliding to
nucleons due to smaller differences in the temperatures of the two
species.


\begin{thebibliography}{99}	
\bibitem{Casanellas:2009dp} 
  J.~Casanellas and I.~Lopes,
  Astrophys.\ J.\  {\bf 705}, 135 (2009)
  [arXiv:0909.1971 [astro-ph.CO]].
 
\bibitem{PerezGarcia:2010ap} 
  M.~A.~Perez-Garcia, J.~Silk and J.~R.~Stone,
  Phys.\ Rev.\ Lett.\  {\bf 105}, 141101 (2010)
  [arXiv:1007.1421 [astro-ph.CO]].
  
\bibitem{Casanellas:2010he} 
  J.~Casanellas and I.~Lopes,
  Mon.\ Not.\ Roy.\ Astron.\ Soc.\  {\bf 410}, 535 (2011)
  [arXiv:1008.0646 [astro-ph.CO]].
  
\bibitem{Lopes:2011rx} 
  I.~Lopes, J.~Casanellas and D.~Eugenio,
  Phys.\ Rev.\ D {\bf 83}, 063521 (2011)
  [arXiv:1102.2907 [astro-ph.CO]].

\bibitem{Casanellas:2011qh} 
  J.~Casanellas and I.~Lopes,
  Astrophys.\ J.\  {\bf 733}, L51 (2011)
  [arXiv:1104.5465 [astro-ph.SR]].

\bibitem{Casanellas:2011kf} 
  J.~Casanellas, P.~Pani, I.~Lopes and V.~Cardoso,
  Astrophys.\ J.\  {\bf 745}, 15 (2012)
  [arXiv:1109.0249 [astro-ph.SR]].
                  
\bibitem{PerezGarcia:2011hh} 
  M.~A.~Perez-Garcia and J.~Silk,
  Phys.\ Lett.\ B {\bf 711}, 6 (2012)
  [arXiv:1111.2275 [astro-ph.CO]].
  
\bibitem{PerezGarcia:2011jb} 
  M.~A.~Perez-Garcia, J.~Silk and J.~R.~Stone,
  AIP Conf.\ Proc.\  {\bf 1441}, 525 (2012)
  [arXiv:1108.5206 [astro-ph.CO]].
 
\bibitem{Brayeur:2011yw} 
  L.~Brayeur and P.~Tinyakov,
  Phys.\ Rev.\ Lett.\  {\bf 109}, 061301 (2012)
  [arXiv:1111.3205 [astro-ph.CO]].
  
\bibitem{Iocco:2012wk} 
  F.~Iocco, M.~Taoso, F.~Leclercq and G.~Meynet,
  Phys.\ Rev.\ Lett.\  {\bf 108}, 061301 (2012)
  [arXiv:1201.5387 [astro-ph.SR]].

\bibitem{Horowitz:2012jd} 
  C.~J.~Horowitz,
  arXiv:1205.3541 [astro-ph.HE].
  
\bibitem{Capela:2012jz} 
  F.~Capela, M.~Pshirkov and P.~Tinyakov,
  arXiv:1209.6021 [astro-ph.CO].
      
\bibitem{Lopes:2012af} 
  I.~Lopes and J.~Silk,
  Astrophys.\ J.\  {\bf 757}, 130 (2012)
  [arXiv:1209.3631 [astro-ph.SR]].
 
\bibitem{Goldman:1989nd}
  I.~Goldman and S.~Nussinov,
  Phys.\ Rev.\  D {\bf 40}, 3221 (1989).
  
\bibitem{Kouvaris:2007ay}
  C.~Kouvaris,
  Phys.\ Rev.\  D {\bf 77}, 023006 (2008)
  [arXiv:0708.2362 [astro-ph]].
  
\bibitem{Bertone:2007ae}
  G.~Bertone and M.~Fairbairn,
  Phys.\ Rev.\  D {\bf 77}, 043515 (2008)
  [arXiv:0709.1485 [astro-ph]].
  
\bibitem{Sandin:2008db} 
  F.~Sandin and P.~Ciarcelluti,
  Astropart.\ Phys.\  {\bf 32}, 278 (2009)
  [arXiv:0809.2942 [astro-ph]].

\bibitem{McCullough:2010ai} 
  M.~McCullough and M.~Fairbairn,
  Phys.\ Rev.\ D {\bf 81}, 083520 (2010)
  [arXiv:1001.2737 [hep-ph]].

\bibitem{Kouvaris:2010vv} 
  C.~Kouvaris and P.~Tinyakov,
  Phys.\ Rev.\ D {\bf 82}, 063531 (2010)
  [arXiv:1004.0586 [astro-ph.GA]].

\bibitem{deLavallaz:2010wp}
  A.~de Lavallaz and M.~Fairbairn,
  Phys.\ Rev.\  D {\bf 81}, 123521 (2010)
  [arXiv:1004.0629 [astro-ph.GA]].
    
\bibitem{Kouvaris:2010jy} 
  C.~Kouvaris and P.~Tinyakov,
  Phys.\ Rev.\ D {\bf 83}, 083512 (2011)
  [arXiv:1012.2039 [astro-ph.HE]].

\bibitem{Ciarcelluti:2010ji}
  P.~Ciarcelluti and F.~Sandin,
  Phys.\ Lett.\  B {\bf 695}, 19 (2011)
  [arXiv:1005.0857 [astro-ph.HE]].
 
\bibitem{Kouvaris:2011fi} 
  C.~Kouvaris and P.~Tinyakov,
  Phys.\ Rev.\ Lett.\  {\bf 107}, 091301 (2011)
  [arXiv:1104.0382 [astro-ph.CO]].
 
\bibitem{McDermott:2011jp} 
  S.~D.~McDermott, H.~-B.~Yu and K.~M.~Zurek,
  Phys.\ Rev.\ D {\bf 85}, 023519 (2012)
  [arXiv:1103.5472 [hep-ph]].
  
\bibitem{Kouvaris:2011gb} 
  C.~Kouvaris,
  Phys.\ Rev.\ Lett.\  {\bf 108}, 191301 (2012)
  [arXiv:1111.4364 [astro-ph.CO]].
  
\bibitem{Guver:2012ba} 
  T.~Guver, A.~E.~Erkoca, M.~H.~Reno and I.~Sarcevic,
  arXiv:1201.2400 [hep-ph].
    
\bibitem{Fan:2012qy} 
  Y.~-z.~Fan, R.~-z.~Yang and J.~Chang,
  arXiv:1204.2564 [astro-ph.HE].
  
  \bibitem{Mielke:2000mh} 
  E.~W.~Mielke and F.~E.~Schunck,
  Nucl.\ Phys.\ B {\bf 564}, 185 (2000)
  [gr-qc/0001061].
    
\bibitem{Stoof:1992zz} 
  H.~T.~C.~Stoof,
  Phys.\ Rev.\ A {\bf 45}, 8398 (1992).


\bibitem{Nov}
Ya.B. Zeldovich and I.D. Novikov in Relativistic Astrophysics,vol. 1,
 The University of Chicago Press, Chicago,1972

\bibitem{Yakovlev:2004yr} 
  D.~G.~Yakovlev, O.~Y.~Gnedin, M.~E.~Gusakov, A.~D.~Kaminker, K.~P.~Levenfish and A.~Y.~Potekhin,
  Nucl.\ Phys.\ A {\bf 752}, 590 (2005)
  [astro-ph/0409751].

\bibitem{Lattimer:1994}
J.~M. Lattimer, K.~A. {van Riper}, M. Prakash and M. Prakash,
Astrophys.\ J.\  {\bf 425}, 802 (1994).

\end{thebibliography}
\end{document}